# Quarantine Fatigue: first-ever decrease in social distancing measures after the COVID-19 outbreak before reopening United States

Jun Zhao[1], Minha Lee[1], Sepehr Ghader[1], Hannah Younes[1], Aref Darzi[1], Chenfeng Xiong[1], Lei Zhang[1,*]

[1] Maryland Transportation Institute, Department of Civil and Environmental Engineering, University of Maryland, College Park, Maryland, USA

[*] Herbert Rabin Distinguished Professor (Corresponding Author). Email: lei@umd.edu

**Author Contributions**

S. G., C. X., L. Z.: designed the study; S. G., A. D.: preprocessed the data; J. Z., M. L.: analyzed the data and visualized the results; H. Y.: reviewed literature and wrote the introduction; J. Z., M. L., S. G., A. D., H. Y.: contributed the first draft; All authors helped with the final version and proof-reading.

# Abstract

By the emergence of the novel coronavirus disease (COVID-19) in Wuhan, China, and its rapid outbreak worldwide, the infectious illness has changed our everyday travel patterns. In this research, our team investigated the changes in the daily mobility pattern of people during the pandemic by utilizing an integrated data panel. To incorporate various aspects of human mobility, the team focused on the Social Distancing Index (SDI) which was calculated based on five basic mobility measures. The SDI patterns showed a plateau stage in the beginning of April that lasted for about two weeks. This phenomenon then followed by a universal decline of SDI, increased number of trips and reduction in percentage of people staying at home. We called the observation Quarantine Fatigue. The Rate of Change (ROC) method was employed to trace back the start date of quarantine fatigue which was indicated to be April 15th. Our analysis showed that despite the existence of state-to-state variations, most states started experiencing a quarantine fatigue phenomenon during the same period. This observation became more important by knowing that none of the states had officially announced the reopening until late April showing that people decided to loosen up their social distancing practices before the official reopening announcement. Moreover, our analysis indicated that official reopening led to a rapid decline in SDI, raising the concern of a second wave of outbreak. The synchronized trend among states also emphasizes the importance of a more nationwide decision-making attitude for the future as the condition of each state depends on the nationwide behavior.

# Introduction

The coronavirus disease 2019 (COVID-19) pandemic is one of the worst global health crises seen in decades. Initially a regional phenomenon, COVID-19 has spread through the entire world in the matter of months and has prompted drastic international and national measures. The United States has by far the most confirmed cases and deaths in the world, with over 1.99 million cases and more than 112,000 confirmed deaths as of June 8th 2020 (1).

On March 13th, 2020, the U.S. government announced a national state of emergency, prompting state governments to begin implementing emergency containment measures (2). On March 19th, 2020, California became the first state to implement a "Stay-at-home" order (3). To preserve the public health and safety (4) and prevent the healthcare system from being overwhelmed (5), this containment measure aims to limit contact with other people by closing non-essential businesses, banning gatherings larger than 10 people, instituting a minimum distance between people, and requiring self-isolating when COVID-19 symptoms arise (6). By the end of April, all but eight states had implemented some form of a "Stay-at-home" order (7).

The impact of the rapid spread of COVID-19 and the government orders prompted a significant reduction in human mobility in the U.S. By Mid-March, daily per person miles traveled had dropped by over 20% from the benchmark days in early February. In the week after government orders were issued, per person miles traveled reduced by an additional 10.8% nationally. While this reduction initially appeared promising, a new phenomenon occurred beginning in Mid-April in which Americans were observed to stay home less. This trend, decline in social distancing and increase in traveling, indicates that Americans are abiding less to the mandatory "Stay-at-home" orders nationally. A combination of warmer weather, tiredness of staying at home and unaffordability of living while unemployed is likely to have initiated this "Quarantine Fatigue" (8, 9). This new relationship is alarming in that less social distancing could easily prompt a second wave of COVID-19 cases, not only in states that are reopening, but also in states that have been strict about not reopening anytime soon at a large scale.

Staying at home is closely related to quarantining. Over 300 million people in the U.S. were ordered to remain at home. While aiming at reducing the spread of a pandemic, self-isolating can have negative effects on mental health (10–12) and can be particularly detrimental to those who live alone and elderly people (12). Thus, literature in the past has focused on the balance between the effects of self-isolation on mental health and public health (13). The current situation is, however, unprecedented in our lifetime and thus, the literature is beginning to expand to other fields. Containment measures have shown to be effective in the context of other infectious diseases (14) and given the scale of the COVID-19 pandemic, preventing the spread is paramount to not only reducing the number of deaths related to COVID-19 but not overwhelming the healthcare system in treating other illnesses (5). In a recent literature review, (13) concluded that information is key - people in quarantine need to understand the gravity of the situation. They also noted that voluntary quarantine is associated with less distress and fewer complications; but it is not clear from the study whether it accelerates the number of cases as compared to strict quarantine. With the data of human mobility during the pandemic, the daily confirmed cases and the date each state started the shelter-at-home orders, we are able to explore their relationship.

# Results

## Nationwide: social distancing inertia and quarantine fatigue

Our research team at the University of Maryland has explored the multifaceted impacts of COVID-19 during the unprecedented pandemic in the U.S. We have built a daily updated platform regarding the economy, sociodemographic groups, and healthcare system trends associated with the virus spread and mobility tendency (15). While the data is available in the COVID-19 Impact Analysis Platform (https://data.covid.umd.edu/), we mainly examine the social distancing trend with mobility measures in this paper. Our study analyzes weekdays from January 6th to May 1st with the measures smoothed out by a five-day moving average to reduce day-to-day noise.

We demonstrate the mobility trends that present social distancing inertia and quarantine fatigue in two folds: 1) nationwide mobility trends are presented over time by utilizing the social distancing index, the percentage of people staying home, number of work trips per person, number of non-work trips per person, and trip distance; and 2) a social distancing index (SDI) is utilized to portray human behavior during the pandemic. SDI, developed by the research team, is chosen to examine the trends as no single mobility metric can sufficiently capture human mobility changes. SDI is a score-based index that measures the extent of social distancing practices in a geographical area by considering the behavior of the residents and visitors of the area simultaneously. The index is calculated at the state and county levels based on the mobility metrics generated from mobile device location data. For each area, a score between zero to one hundred is assigned, where zero denotes no social distancing practices and one hundred indicates perfect social distancing in comparison to benchmark days prior to the COVID-19 pandemic. The five metrics included in the calculation of SDI were percentage of residents staying home, daily work trips per person, daily non-work trips per person, distances traveled per person, and percentage of out-of-county trips. To consider the importance of each variable properly in the weighting procedure, both real-world observations and conceptual guidelines are taken into the account. A more detailed description of the SDI functional form can be found in our earlier work (16, 17).

Figure 1 displays the mobility tendency associated with COVID-19 cases in the U.S. The national emergency proclamation on March 13th is marked by the black dotted vertical line. Almost immediately after the national emergency declaration, a large number of people started sheltering at home and reducing their daily movements. From mid-March to late March, the nationwide SDI increased sharply from 15 to around 50 within 10 days. However, from early April to mid-April, the SDI reached a plateau and stopped increasing. We call this phenomenon Social Distancing Inertia (18). After mid-April, we observed a downfall of the social distancing measures even though no states had eased the mobility restriction yet. From mid-April to early May, the percentage of people staying at home decreased from 34% to 31.5%. In addition, the average number of trips per person per day inclined from 2.75 to 3.0; the non-work trip rates increased from 2.3 to 2.5; and per-person daily mile traveled increased from 23.6 to 24.7. From these changes in mobility metrics, one can note that people tend to leave their home more frequently and travel further (17). We refer to this phenomenon as Quarantine Fatigue and this period is shaded in Figure 1.

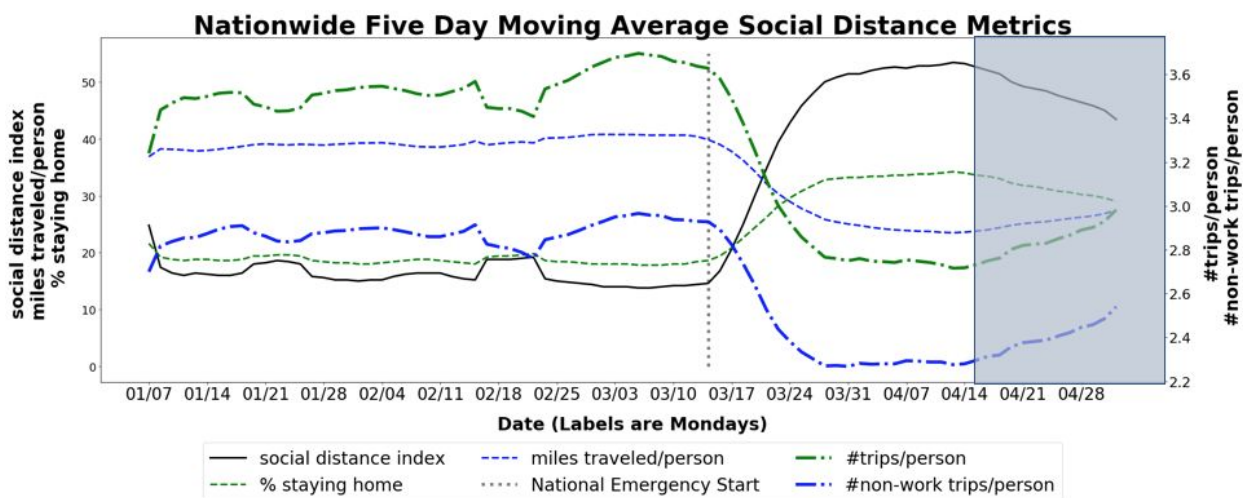

**Figure 1**: Nationwide five day moving average social distancing metrics

## Quarantine fatigue starts before reopening

To better understand Quarantine Fatigue, we next explore when the nation starts to experience the fatigue period by checking the momentum of SDI. Momentum is literally a measurement of mass in an object's movement in physics, while it also refers to the speed of price changes over time to understand a trend in a stock market. We adapt this concept of momentum to measure the social distancing inertia and quarantine fatigue. Among numerous types of momentum oscillators; the rate of change (ROC) is applied here, which is the classical, yet effective and intuitive centered oscillator (19).

This paper defines ROC as the percentage rate of change in SDI over a period and compares its fluctuations above and below zero. The bigger the SDI difference between the current and predetermined previous day, the higher the value of the ROC oscillator. When the indicator is above 0, the percentage SDI change is positive. When the indicator is below 0, the percentage change is negative. We define 'inertia' once a deep elbow pattern of ROC is observed, while 'fatigue' is defined by days with constant negative ROCs.

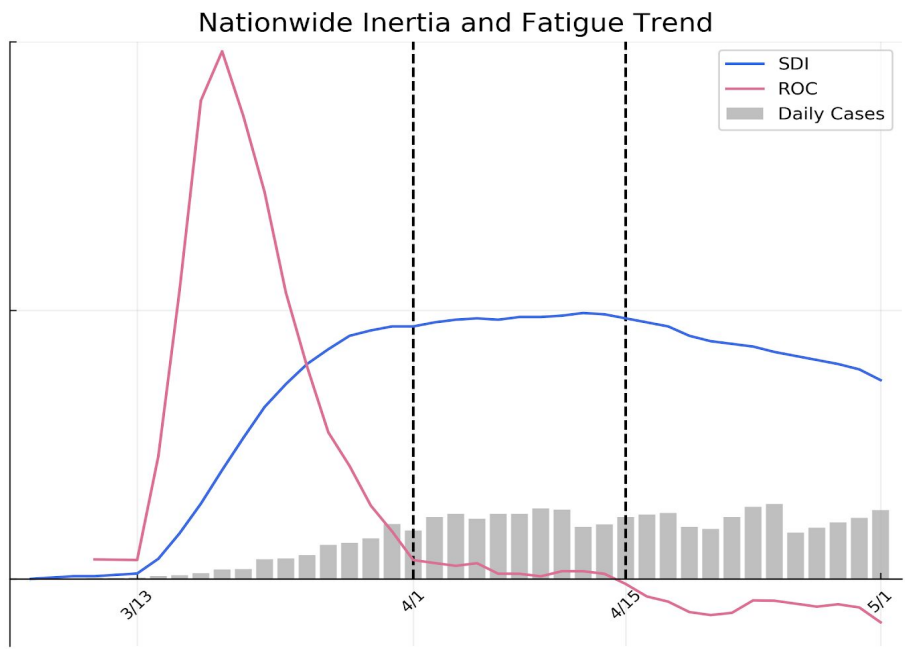

**Figure 2**: Nationwide ROC curve

In Figure 2, after the declaration of the national emergency on March 13th, the ROC curve of SDI increased sharply and remained above zero. The ROC curve then dropped down to nearly zero on April 1st and stayed at a low level until April 15th. After April 15th, the ROC falls below zero which indicates that the SDI is continuously decreasing. The period between April 1st to April 15th is defined as Social Distancing Inertia and April 15th to April 30th as Quarantine Fatigue. To check whether the decline is significant or not we have conducted a t-test with H0: the social distancing index is smaller in the week after April 15th compared with the social distancing index of the week before April 15th. The p-value equals 2.4e-05 for failing to reject H0.

The nationwide curve demonstrates that after remaining in a high-level conformity of stay-at-home orders, Americans are gradually going outside. As over forty states are opening up again starting from the end of April, it is reasonable to assume that the SDI will continue to drop. Even though there must be comprehensive factors that determine reopening policies or the increased mobility observed, this phenomenon might put people into the risk of infection by only focusing on the effect of population migration flow on COVID-19 (20).

## State level fatigue universality

### Universal decline of SDI

Besides the nationwide quarantine fatigue, another observation to mention is the universality of the fatigue among all states. Even though almost all the states are still in the outbreak, the SDI has decreased by 6.5% on average. Figure 3 shows the average SDI trend together with the number of confirmed cases per thousand population in all the states.

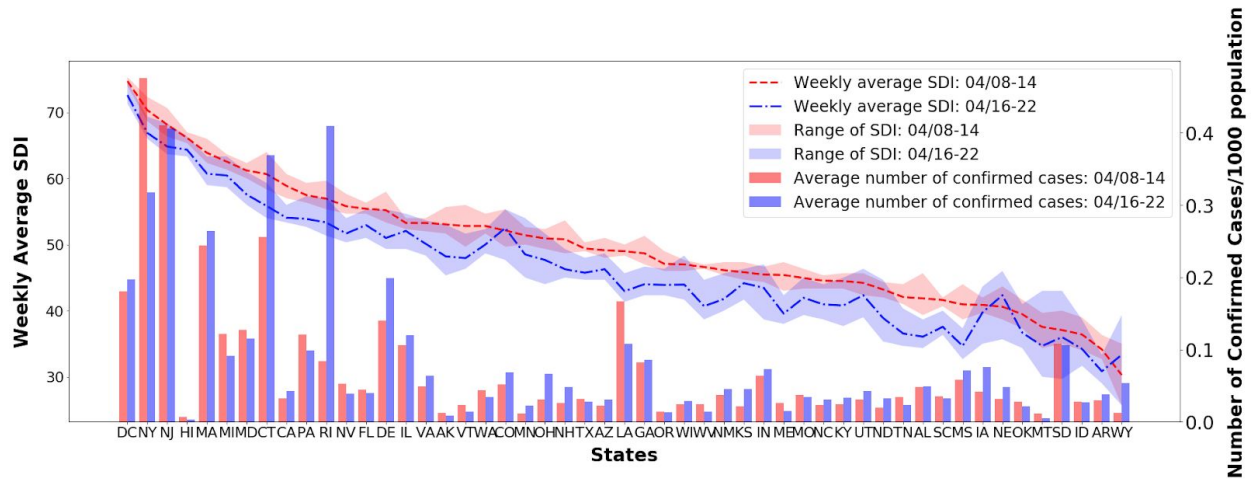

**Figure 3**: SDI trend: ordered by SDI of the week before fatigue starts

The red line is the average SDI for the week before April 15th and the blue line is the average SDI for the week after April 15th. The average SDI decline is universal among states except Colorado, Nebraska and Wyoming. It reflects that almost all the states had experienced an increase of public mobilities after mid-April. In addition, the variance of the week before the fatigue (the red shade) is smaller than the week after (the blue shade). It may indicate that mobility behavior has become more dynamic compared to the earlier pandemic stage and more contradictory against the government-issued social distancing urge. The higher SDI states tend to have the smaller variance. In that, the states with a higher SDI, such as the District of Columbia, New York and New Jersey, experience relatively slight decline of SDI and people tend to maintain their daily travel behaviors during these periods. The bar chart with two colors for each state demonstrates the number of average confirmed cases per thousand population of the weeks before and after April 15th, respectively. States with a higher increase of cases per population like Connecticut and Rhode Island still experience the decrease of SDI corresponding to the nationwide fatigue trend, rather than increasing in-state trends of new cases. This trend supports the existence of nationwide generality of quarantine fatigue.

## State level comparison

Since the universal decrease of SDI is observed, we further explore the starting date of quarantine fatigue for each state and the relationship between the SDI and the reopening dates for the partially reopening states by May 1st.

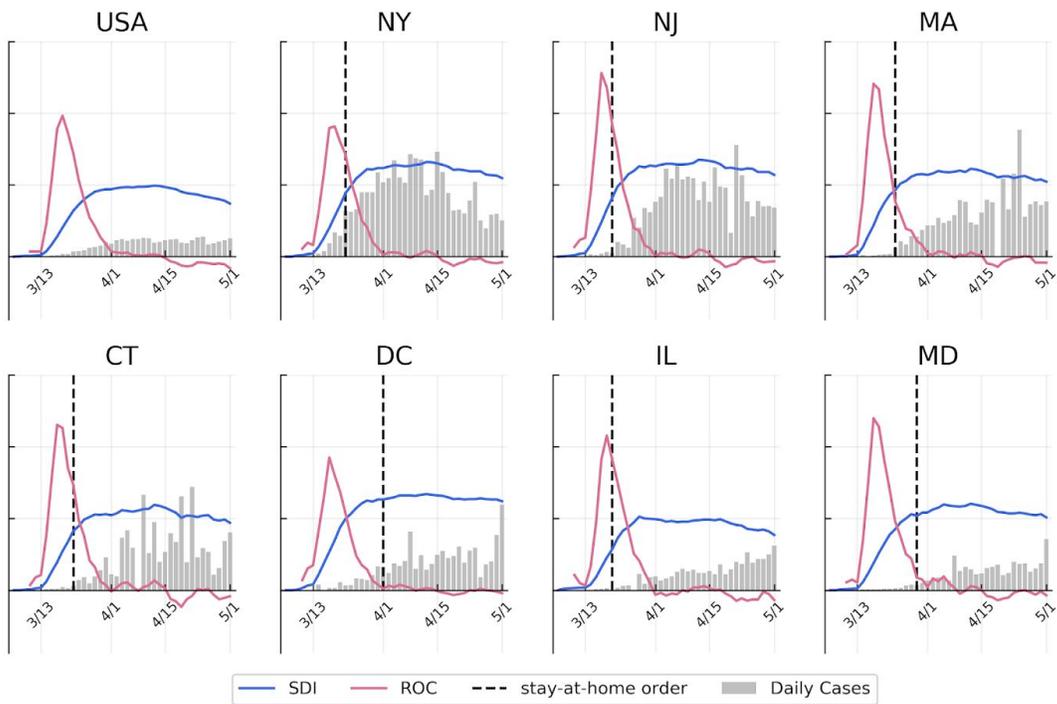

**Figure 4a**: State level ROC curve ordered by confirmed number per population

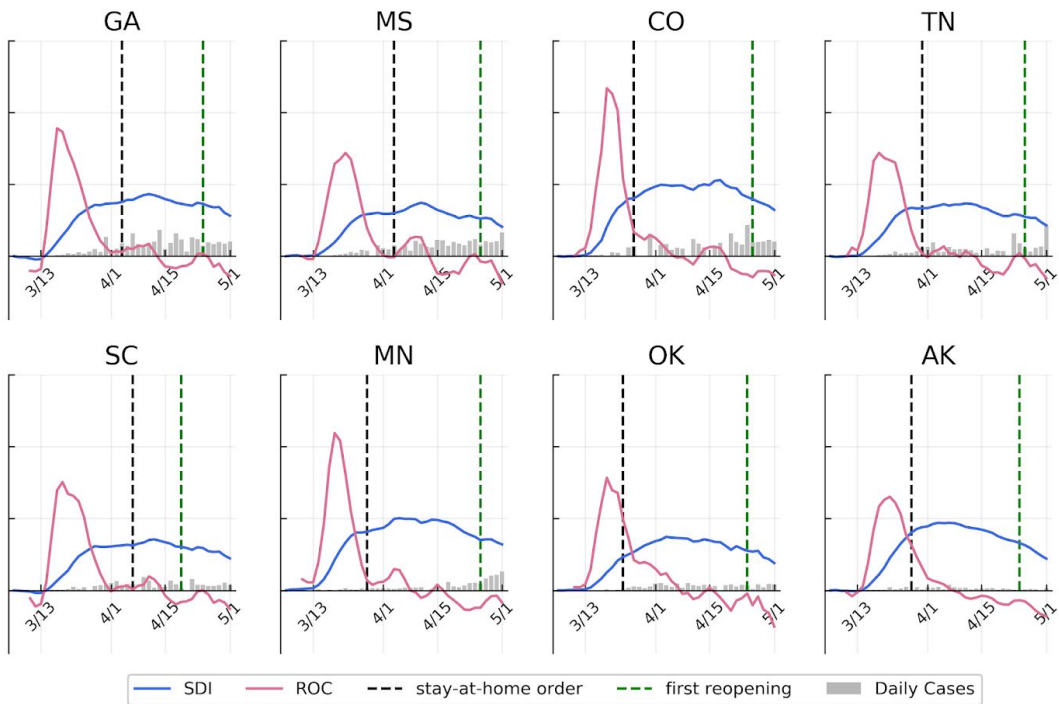

**Figure 4b**: State level ROC curve ordered by start date of reopening

Figure 4a illustrates the SDI and its ROC for the nation and seven highly affected states by the virus. Three states at the upper row are those with the highest cumulative COVID-19 confirmed cases per population as of May 1st, 2020. The bottom row illustrates states with increasing daily trends of confirmed cases. The social distancing had become active immediately after the national emergency declaration on March 13th, even though the scale varies among states. Afterwards, all the states entered the inertia stage after about two weeks of rapid increase. Then, staying in the inertia for another two weeks, the deepest elbow of ROC occurs around April 15th, indicating the start of fatigue.

Regardless of the in-state confirmed cases or stay-at-home order date, all seven states have presented the immediate increase of the SDI curve as of March 13th. The possible reason is that the impact of the national emergency declaration and the rapid surge of the nationwide daily confirmed new cases might have led to the population behavior. Another point to note is the general duration of each phenomenon: the rapid increase of social distancing is simultaneously observed during the time from March 13th to April 1st, the inertia from April 1st to April 15th and the fatigue after April 15th. This observation also supports the universality in Figure 3. The SDI curve is either along the way to the highest peak at the date of stay-at-home orders or already reached to the peak in advance, which indicates the order might have not been the major factor for active reaction to the virus. Those four states with the increasing COVID-19 trend at the bottom row (Connecticut, District of Columbia, Illinois and Maryland) also present both inertia and fatigue during the same time window. The state-wide social distancing tendency corresponds with the nationwide trend, which indicates that there exist people who are behaving in response to the nationwide pandemic status not only to the state level condition of COVID-19.

States moving into a new phase of reopening business also demonstrate a similar but more fluctuating pattern (Figure 4b). Figure 4b illustrates eight states where partial reopening is initiated before May 1 and they are relatively at lower risk in terms of COVID-19 cases. The fatigue stages are observed before easing of mobility restrictions indicating that the general public left their houses and went outside even before states had lifted the stay-at-home order. In general, the ROC curve presents more fluctuation with more round-shaped peaks compared to the higher risk regions. Moreover, in states like Georgia, Mississippi, Tennessee and Oklahoma, the fatigue was slowing down with the ROC reaching near the center line at zero. However, after the declaration of reopening, the rate of change began to drop even faster, putting more people in the risk of infection. Lastly, future observations may be required to explore the potential impact of ease of mobility restriction on the COVID-19 severity. All 15 states in Figure 4 reciprocate between inertia and fatigue periods and the fatigue trend tends to be more obvious among lower risk states with reopening.

# Discussion

In the first two weeks after declaring the national emergency, the social distancing metrics had changed significantly: more people stayed at home, fewer trips were observed, and travel distance became shorter. However, by the third week of March, the trend of all metrics on social distancing had reached a plateau and remained steadily high for two weeks. We call this phenomenon 'Social Distancing Inertia' (18). Starting mid-April, the national SDI began to decline. We refer to the phenomenon of shifting from self-restriction to pre-pandemic behavior as "Quarantine Fatigue".

To check the universality of this increase of physical activities in different states, we have investigated the ROC curve of five day moving average SDI and also the average SDI for the week before and after April 15. Despite the different changes in the confirmed cases per thousand population of each state, we could find a universal decline of SDI for almost all the states in the nation. It shows that people reacted to the nationwide situation of the pandemic not only to the state level conditions.

The national quarantine fatigue started around April 15th, before any state had announced reopening policies while the first state to announce reopening was South Carolina by easing restrictions on outdoor and recreation, retail and beauty on April 20th. The universal decline of SDI indicates that on a nationwide level people are going outside more without any states declaring the re-openness. There could be multiple reasons for this phenomenon. One may assume that it is because households must start working on site to pay for the bills and these are essential trips. However, by digging deeper into the data, we could find out more evidence to support the phenomenon of fatigue. After mid-April, the total trips traveled per person increased from 2.75 to 3.0 as seen in figure 3. Among the 0.25 increase, non-work trips have counted for 80%, changing from 2.3 to 2.5. Therefore, only 20% of the increased trips are work trips and most of the increased travels are non-work trips which don't support the assumption. One may also argue that the fatigue could relate to the upcoming protests over lockdown for states like Michigan (21). It was true that Michigan residents started the protest on April 15th which coincidentally matched with the start date of the quarantine fatigue nationwide. However, it actually supports our finding that without enforcement, or consistent and effective knowledge propagation of the infectious disease, people tend to get away from the self-isolated voluntary quarantine situation. They would involve themselves in more outdoor activities no matter whether a reopening is announced in the state or not.

# Data Description

## Mobile device location data

In this study, we used mobile device location data from several leading location data providers. Our platform utilizes location data from more than 100 million anonymized monthly active users (MAU) in the United States. Generally, mobile device location data is continuously being collected from various technologies such as cellphone towers, GPS, and location-based services (LBS) (22–25). Our data is from LBS of several providers. The data contains information about latitude and longitude coordinates, timestamp, and a measure of accuracy (26). Figure 5 shows the data cleaning procedure adapted and the details can be found in our previous work (15, 18), which addresses four dimensions of data quality assessment: consistency, accuracy, completeness, and timeliness (27).

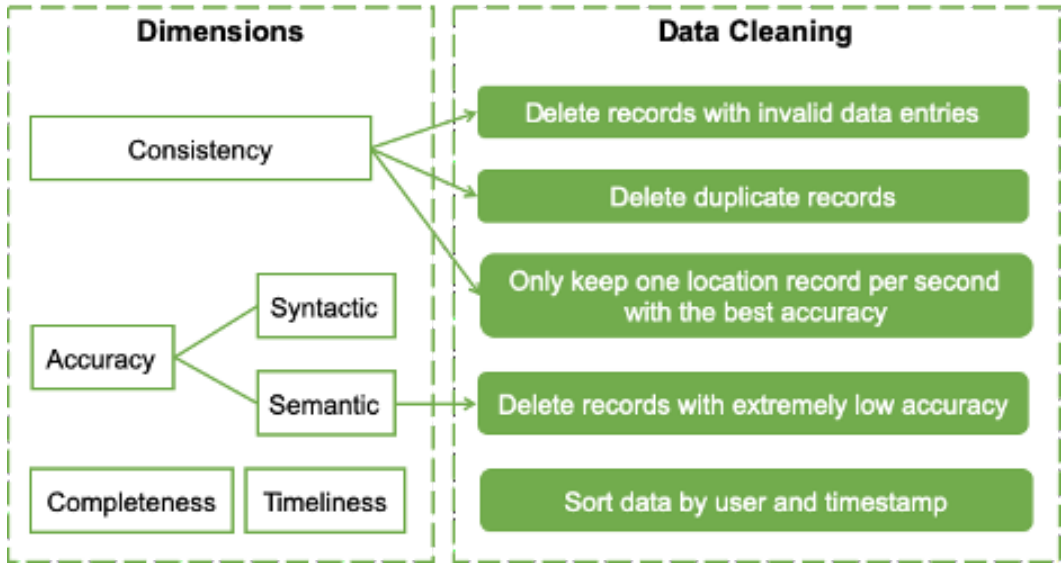

**Figure 5.** State-of-the-practice data cleaning procedure (15)

## COVID19 case data

John Hopkins University's Center for System Science and Engineering has created an interactive online dashboard (1) that presents and visualizes information about COVID19, in terms of confirmed cases, confirmed deaths, and recoveries. They also provide free access to the data through a GitHub repository (28) where the U.S. data is aggregated to the both state and county level. We have utilized this dataset to integrate our mobility statistics with COVID19 case information. The data statistics are temporally aggregated to daily values. All statistics are also spatially aggregated to county-level, state-level, and national-level metrics for privacy protection. Counties are the smallest unit of analysis on our platform.

## Data availability

All the aggregated metrics used on platform are available to general public on http://data.covid.umd.edu/. Access to the anonymized location data is restricted to the Maryland Transportation Institute. The new confirmed case data is available at https://github.com/CSSEGISandData/COVID-19/tree/master/csse_covid_19_data.

# Materials and Methods

The first step of the methodology is data cleaning followed by the clustering of location observations into activity locations and identifying home and work census block groups (CBG). We examine both the temporal and spatial distribution of activity locations to identify home and work CBGs. Next, we apply a trip identification algorithm that evaluates which location points form a trip together and identifies trip origin, trip destination, departure time, and arrival time. Then, a multi-level weighting procedure is utilized to expand our sample to population and provide population-level statistics. The methodologies have been previously developed and validated based on various independent datasets such as National Household Travel Survey (NHTS), American Community Survey (ACS), and longitudinal employer-household dynamics (LEHD), and peer-reviewed by an external experts panel in a U.S. Department of Transportation Federal Highway Administration's Exploratory Advanced Research Program project, titled "Data analytics and modeling methods for tracking and predicting origin-destination travel trends based on mobile device data" (29). Afterward, we integrate the mobility metrics with population and COVID-19 case data to produce the metrics available on the platform. Figure 6 shows a summary of the methodology.

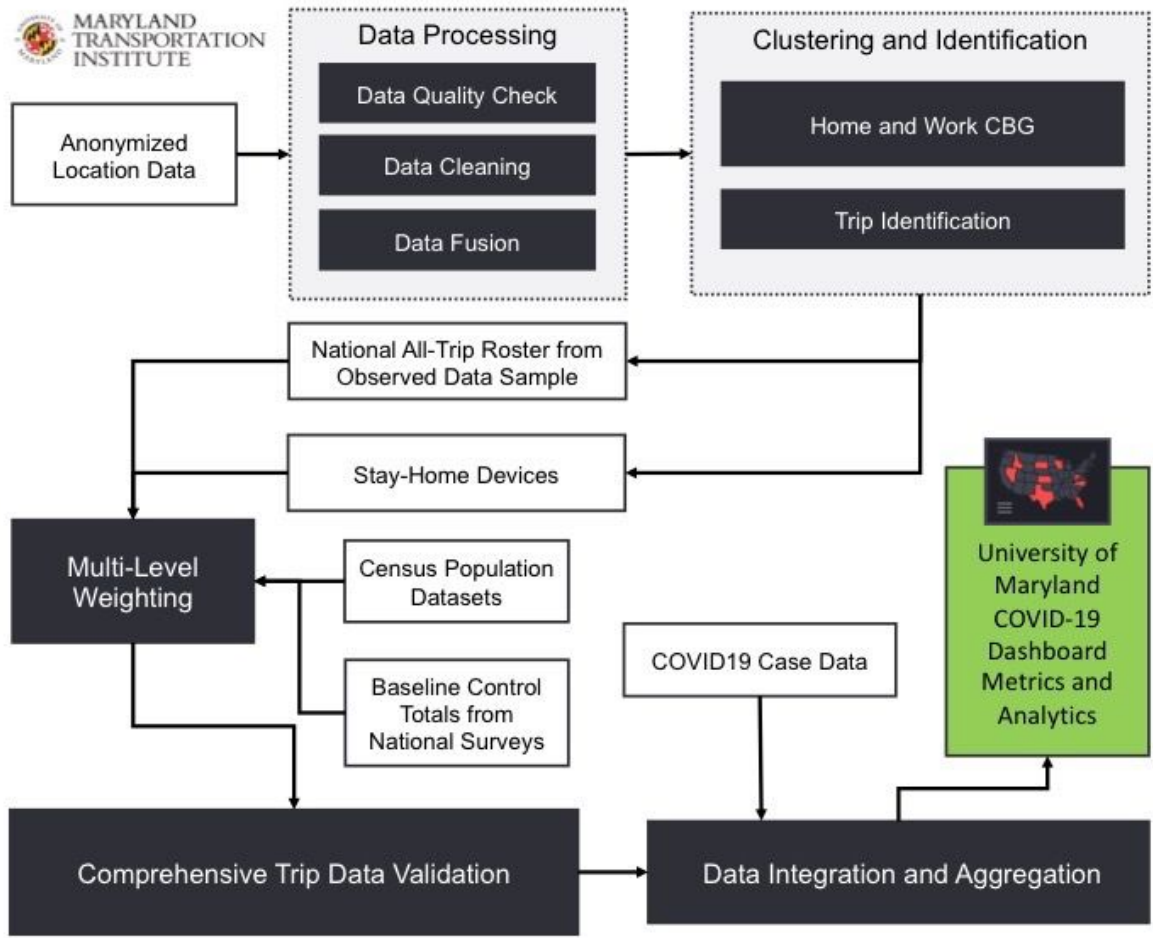

**Figure 6.** Methodology (15)

## Trip identification

Trips are the unit of analysis in our study. Mobile device location data does not include trip information. Location sightings are continuously being generated while the device moves, stops, stays static, or starts a new trip. As a result, we developed a trip identification algorithm, which can detect the location sightings that form a trip together.

We first sort all device observations by time. The algorithm assigns a random ID to each trip it identifies. The algorithm assigns "0" as the trip ID of these locations to tag them as static points. For every location point, we calculate distance, time, and speed between the point and its immediate previous and next points, if they exist. Three hyperparameters need to be set for the algorithm: distance threshold, time threshold, and speed threshold. The speed threshold is used to identify if a location point is recorded on the move. The distance and time threshold are used to identify stay locations and trip ends.

Next the recursive algorithm checks every point to identify if they belong to the same trip as their previous point. If they do, they are assigned the same trip ID. If they do not, they are either assigned a new hashed trip id when their *speed from > speed threshold* or their trip ID is set to "0" when their *speed from < speed threshold*. Identifying if a point belongs to the same trip as its previous point is based on the point's "speed to", "distance to" and "time to" attributes. If a device is seen in a point with *distance to > distance threshold* but is not observed to move there *speed to < speed threshold*, the point does not belong to the same trip as its previous point. When the device is on the move at a point where *speed from >= speed threshold*, the point belongs to the same trip as its previous point; but when the device stops, the algorithm checks the radius and dwell time to identify if the previous trip has ended. If the device stays at the stop (points should be closer than the distance threshold) for a period of time shorter than the time threshold, the points still belong to the previous trip. When the dwell time reaches above the time threshold, the trip ends, and the next points no longer belong to the same trip. The algorithm does this by updating "time from" to be measured from the first observation in the stop, not the point's previous point. The algorithm may identify a local movement as a trip if the device moves within a stay location. To filter out such trips, all trips that are shorter than 300 meters are removed.

## Activity identification

Next, we use spatial and temporal distribution of activity locations to identify the home and work census block groups (CBG). The first step is activity clustering. We have applied HDBSCAN (30) clustering algorithm to identify activity locations with device observations. Utilizing the cleaned multi-day location data as input, we apply an iterative algorithm until no cluster has a radius larger than two miles. The iterative algorithm consists of two parts: HDBSCAN based on a minimum number of point parameters and filtering non-static clusters based on time and speed checks. In case of splitting a single activity, the method combines nearby clusters after finalizing the potential stay clusters.

The next step is home and work CBG identification. The methodology applied here is to identify the most frequently visited home and work clusters. The framework examines both temporal and spatial features for the entire activity location list instead of a fixed simply setting fixed time period for each type. In this way, it could capture more diverse work schedules and simultaneously detect the employment type for each device.

## Weighting

The last step is weighting the sample data to produce the statistics for the whole population. We need to not only expand the sample to the population but also adjust the trip rate on a personal level as we may not necessarily capture all trips of an observed device. We actually applied a simple weighting method in order to have a timely analysis, namely county-level device weights and state-level trip weights. For county level device weight, we simply count the number of devices included in our sample for each county and calculate the ratio between the population of the county and the number of devices observed in our sample. For state-level trip weights, we use the first two weeks of February in our sample to calculate average trip rate (trips/person) for residents of each state.

## Acknowledgments

We would like to thank and acknowledge our partners and data sources in this effort: (1) Amazon Web Service and its Senior Solutions Architect, Jianjun Xu, for providing cloud computing and technical support; (2) computational algorithms developed and validated in a previous USDOT Federal Highway Administration's Exploratory Advanced Research Program project; (3) mobile device location data provider partners; and (4) partial financial support from the U.S. Department of Transportation's Bureau of Transportation Statistics.